\documentclass[%
aip,
amsmath,amssymb,
reprint,%
]{revtex4-1}

\usepackage{graphicx}
\usepackage{dcolumn}
\usepackage{bm}

\usepackage[T1]{fontenc}
\usepackage{mathptmx}
\usepackage{etoolbox}
\usepackage{xcolor} 
\usepackage{soul}

\makeatletter
\def\@email#1#2{%
	\endgroup
	\patchcmd{\titleblock@produce}
	{\frontmatter@RRAPformat}
	{\frontmatter@RRAPformat{\produce@RRAP{*#1\href{mailto:#2}{#2}}}\frontmatter@RRAPformat}
	{}{}
}%
\makeatother
\begin{document}
	
	\preprint{AIP/123-QED}
	
	\title[Bimodality of local structural ordering]
{Bimodality of local structural ordering in extremely confined  hard disks
}
	\author{A. Trokhymchuk}
	\email{andrij.trokhymchuk@fkkt.uni-lj.si}
		\altaffiliation[Also at: ]{ Department of Soft Matter Theory, Institute for Condensed Matter Physics of the National Academy of Sciences of Ukraine, Lviv, Ukraine  
	}
	\affiliation{Faculty of Chemistry and Chemical Technology, University of Ljubljana, Ljubljana, Slovenia
}
	\author{A. Huerta}
	\altaffiliation[On sabbatical leave at: ]{Departamento de Fisica Quimica, Instituto de Fisica, Universidad Nacional Autonoma de Mexico (UNAM), 
Mexico, Distrito Federal, Mexico 
}
	\affiliation{Facultad de Fisica, Universidad Veracruzana, Xalapa, Mexico
	}%
	
	\author{T. Bryk}
	\affiliation{Institute for Condensed Matter Physics of the National Academy of Sciences of Ukraine, Lviv, Ukraine
}
	\affiliation{Institute of Applied Mathematics and Fundamental Sciences, Lviv Polytechnic National University, Lviv, Ukraine
}
	
	\date{\today}
	
	\begin{abstract}
	 By combining computer simulations and a unit cell model approach, we study apparent bimodality of local structural ordering in a system of confined hard disks. It is shown that a two-dimensional (2D) array of hard disks being confined laterally within a quasi-1D hard wall channel of the width commensurate the bulk 2D triangular lattice at disk close packing, possesses a bimodal probability distribution for the distance between disk's left and right nearest neighbors. The observed feature aligns with the concept of locally favored structures intensively exploited in the studies of anomalous thermodynamic and kinetic behavior of hydrogen-bonding fluids, except that the reported case is driven by entropic bonding only. The bimodality is observed in a range of densities associated with {the vicinity of freezing transition} in bulk 2D hard disks, indicating a crossover from the "gas-like" to "liquid-like" state in confined quasi-1D hard disks. Such a phenomenon was not reported for a bulk 2D hard disks and is physically unexpected for confined q1D hard disks.
	\end{abstract}
	
	\maketitle
	
	
	\section{\label{sec:level1} Introduction}

The hard spheres and hard disks, as well as the hard-core-based  model systems more generally, in spite of being a rather crude  representation of the real substance  still are capable to  recover a number of its basic properties~\cite{Lowen}. This can be done either directly or via serving as the reference system in various  types of perturbation techniques and others tools of statistical mechanics of fluids~\cite{HansenMcD}. The range of successful applications scans such important areas of research as structure, thermodynamics including phase transitions, and dynamics of liquid state of matter. The reasons why the hard-core-based approaches are so successful in describing liquid phase are not only because of the availability of accurate analytic results for the hard-sphere fluid properties, but primarily because of the flexibility of the hard-core model. This means that in the range of typical liquid densities the model does not exhibit the vapor-liquid phase transition that makes it possible to vary the hard-core diameter (and hence the packing density) over a large range of values 
and tune thus the results to the wished outcome~\cite{adtjsf2010}. The van der Waals excluded volume concept and the role it is playing for the success of microscopic understanding vapor-liquid phase transition is an illustration~\cite{HansenMcD}. 

Nowadays it is well-known that hard-core systems themselves may undergo the phase transitions in the range of liquid densities close to fluid-solid  {transition}~\cite{HSRMP2024,KofkeMonson2000}. So far there are no theories of the dense hard-core systems and these transitions are detected by means of computer simulations exclusively while their peculiarities depend on the dimensionality of the hard-core system under study.  Considering two-dimensional (2D) hard disks,  the freezing-melting transition was discovered by Alder and Wainright~\cite{AlderPR1962} in 1962 for the system of 800 particles. Later it has been confirmed by many others~\cite{Zollweg1992} using systems of up to $\,10^6\,$ hard disks~\cite{Bernard2011} and nowadays it is an accepted phenomenon although there are no full consensus towards its mechanism~\cite{Li2022}. 

{What we are calling "the extremely confined hard disks", represents the 2D hard disks filled into a narrow  2D hard-wall channel of the shortest width still possessing at disk close packing an ordering commensurate  with the bulk 2D triangular lattice.
The most studied of such channels is one with the widths ($1+\sqrt{3}/2$) of disk hard-core diameter. Besides commensurability with the bulk 2D triangular lattice, so defined extreme confinement also does not allow for disks passing each other. Commonly, such} a system setup is also known as the quasi-one-dimensional (q1D) hard-disk system~\cite{Barker1962}, representing a non-trivial deviation from the exactly solved 1D hard-disk system~\cite{Tonks}. In contrast to its bulk 2D counterpart, the former commonly is not expected to have a phase transition, a fact that is causing the special attention to the q1D hard-disk system, i.e., whether it could exhibit one~\cite{adtPRR2020}. Such a fundamental interest in the q1D hard disks stems from the existence for this system of analytical transfer matrix approach for the isobaric partition function~\cite{Wojc,Kofke1993,Varga2011,Varga2013,Montero2023,Montero2024} and since recently also the exact canonical partition function~\cite{PergamJCP2020}, both in the thermodynamic limit. Indeed, the subsequent theoretical studies~\cite{Varga2011,PergamJCP2020} found that transition from the q1D solid-like to the q1D fluid-like state is quite sharp in the density scale. However, none of them indicated any genuine discontinuity in thermodynamic functions~\cite{PergamJCP2020}. Therefore, the main body of theoretical research in this area has been mostly concerned with the efforts to consider these systems as glass formers~\cite{GodfreyPRE2016,YamchiPRE2015}. 
There are also some practical interest because of the possibility to use such a simple model to capture properties of more complex systems~\cite{Mark}
by treating the q1D channel as a pore.

The issue of local structural ordering, which is the main subject of the present study, became an important topic relatively recently since the concept of locally favored structures was introduced to understand the anomalous thermodynamic and kinetic behavior of water-type hydrogen-bonding liquids~\cite{TanakaFD2013,RussoNC2014}. 
It assumes that for a fluid of any complexity, the basic local structure consists of the targeted molecule and its first coordination shell.  In liquids, because of cohesion forces, such a basic local structure extends towards the second coordination shell of the molecules. Upon increasing density (or pressure) and on approaching the freezing transition, the molecules within the first coordination shells undergo local (positional and/or orientational) rearrangement governed by  free energy of entire system seeking the equilibrium state. In the coarse of such a rearrangement the distances between neighboring molecules are changing that affects the local density and the density of entire mesophase.

Hydrogen-bonding liquids and hard-disk fluids are quite different objects. First of all because of the origin of forces between particles, and consequently of different  complexity. Nevertheless, the concept of locally favored structures seems to be the one that is equally important for both. {Independently of the system, the local structure is responsible for local density in the system. The bimodal distribution of the coarse-grained local density was already reported in the studies of the melting transition in the bulk 2D hard-disk system at the  liquid and hexatic phase coexistence}~\cite{Bernard2011,BernardArx2011}.
The essential difficulty in defining locally favored states in the case of complex fluids is to find a proper structural order parameter~\cite{RussoNC2014}. 
In contrast, the relative simplicity of the q1D hard-disk system might allow one to overcome such a difficulty.

Therefore, this paper focuses on a particular q1D hard-disk system of the width commensurate with the triangular lattice of bulk 2D hard disks at close packing. This commensurability allows us to more objectively compare and discuss the behavior and properties of both systems, in particular from the perspective of the presence and mechanism of freezing transition in bulk 2D hard disks. In particular, we found~\cite{adtpre2006} that in such systems under increasing disk density, the assemblies of hard disks are experiencing entropy-driven structural rearrangements towards the formation of local quasiregular hexagons (in bulk 2D hard disks) and commensurable {hexagon's fragments} (in q1D confined hard disks). In both cases, it is accompanied by mutual caging of disks by their nearest neighbors, and each system has a range of densities where caging becomes effective. Based on this observation, in the case of bulk 2D  hard disks, it was argued that nearest-neighbors caging originates the freezing transition, and the triangular unit cell model has been suggested~\cite{adtpre2006}. Following that finding, the q1D implementation of such a unit cell model here is suggested. In this case, it consists of a unique parameter -- the distance between the disk's left and right nearest neighbors.  Moreover, it is shown that this parameter is perfectly suited to serve as the structural order parameter in the entire q1D hard disk system.  The probability distribution of the distance between the disk's left and right nearest neighbors was evaluated by employing computer simulations and its bimodality was found {in the vicinity of densities associated with the freezing  in bulk 2D hard disks. In contrast to bimodal distribution of the coarse-grained local density in the studies of  melting transition  in the bulk 2D hard disks that was used to illustrate the  liquid and hexatic phase coexistence}~\cite{Bernard2011,BernardArx2011}, {the bimodality observed in the present study is an indication of the crossover from the "gas-like" to "liquid-like" state in confined quasi-1D hard disks. The links of this finding to the interpretation of the pair distribution function of the q1D hard disk system at short distances and to the decay of correlations at large distances were discussed as well.}

	\section{Model and Methods} 
	\subsection*{System setup and interaction potentials} 
\begin{figure}[h!]
		\centering
			\includegraphics[scale=0.75]{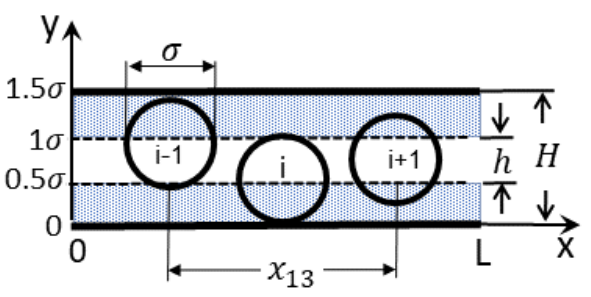}
		\caption{Two-dimensional (2D) hard disks confined at a hard-wall channel of the width $\,H = \sigma + h\,$ with $\,h=(1/2)\sigma\,$.} 
		\label{JCPfig1}
	\end{figure}

The system under study is modeled by $\,N\,$ hard disks of diameter $\,\sigma\,$ placed in a slit pore formed by two horizontal parallel hard lines (2D walls) of length $\,L\,$ that are separated by a distance (pore width) $\,H=\sigma +h\,$  (e.g., see Fig.~\ref{JCPfig1}). The ends of such a hard-wall channel are open. This is a typical setup to study the generic effect of geometric confinement on structural ordering and thermodynamic properties of 2D hard disks~\cite{ChaudhuriJCP2008}.

{There is a set of the  three channel widths that fall into the range from 1 to ($\,1+\sqrt{3}/2\,$) of disk hard-core diameter. Besides the trivial 1D ($\,h=0\,$) array of hard disks,  two other arrays correspond to the horizontal ($\,h=(1/2)\sigma\,$) and vertical ($\,h=(\sqrt{3}/2)\sigma\,$) orientations of 2D triangular lattice that differ by an angle of 30 degrees}~\cite{adtFP2021}. {In} 
the present study, we are interested in the hard disk  confined at a hard wall channel (pore) of the width $\,H\,$ with $\,h=(1/2)\sigma\,$. Such a q1D system of hard disks is necessarily anisotropic~\cite{Kofke1993}. There are two origins of the forces experienced by disks in such a set up. The one originates from disk-disk pair interaction $\,u(r)\,$ given by 
	\begin{equation}
	u(r_{\mathrm{ij}})=\left\{ 
	\begin{array}{ll}
	\infty , & \quad r_{\mathrm{ij}}<\sigma  \\ 
	0, & \quad r_{\mathrm{ij}}\geq \sigma \,.
	\end{array}
	\right.  
	\label{ur}
	\end{equation}
The interaction potential (\ref{ur}) is central, i.e., it is isotropic and $\,r_{\mathrm{ij}}=|\mathbf{r}_{\mathrm{j}}-\mathbf{r}_{i}|\,$ is the distance between disk centers, $\,\sigma\,$ is the hard-disk diameter. The second force experienced by disks is due to impenetrable confining walls which are imposing the disk-wall interaction $\,v(y)\,$ in the form,
	\begin{equation}
	v(y_i)=\left\{ 
	\begin{array}{ll}
	0, & \quad \sigma/2<y_{\mathrm{i}}<h+\sigma/2 \\ 
	\infty , & \quad \mbox{otherwise}\,,%
	\end{array}%
	\right.   \label{vy}
	\end{equation}
i.e., it depends on the normal distance to the wall only. 
	
\subsection*{Unit cell model and structural order parameter}	
\begin{figure}[h!]
 	\centering
 	\includegraphics[scale=0.525]{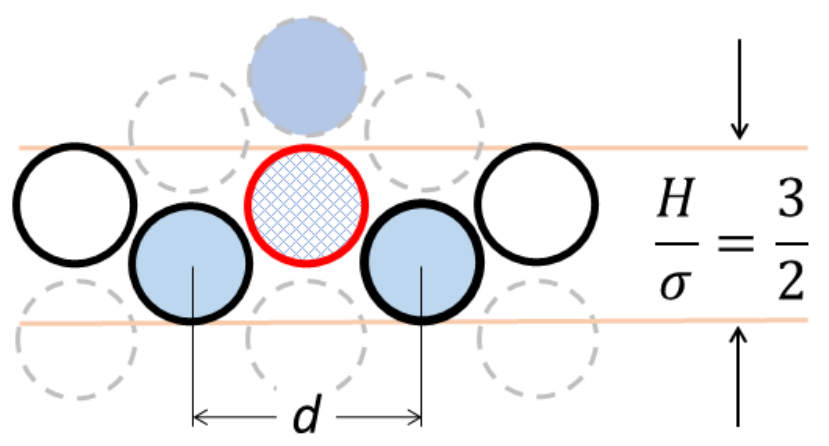}
 	\caption{Schematic ordering in the q1D  hard-disk system of the width $\,H=(3/2)\sigma\,$  at densities before close packing. The three disks filled in blue indicate a triangular unit cell formed by the alternating hexagonal neighbors in a bulk 2D array of hard disks~\cite{adtpre2006}, while {only} the two of them {(rounded with thick solid line)} indicate a unit cell {corresponding} to the considered case of a q1D system of hard disks.} 
 	\label{JCPfig2}
 \end{figure}

Taking into account that the q1D system of hard disks with the width $\,H=\sigma \,$ corresponds to the one-dimensional (1D) case, the variety of q1D systems of hard disks above this width in the range $\,\sigma <H<(1+\sqrt{3}/2)\sigma \,$ could be considered as a bridge connecting 1D and higher dimensions of hard spheres. From this perspective the considered q1D system with width $\,H = (3/2)\sigma\,$ is a special case since at close packing density, it allows the disk ordering commensurate with the bulk 2D triangular lattice (shown schematically in Fig.~\ref{JCPfig2}). Such a commensurability of the considered q1D and the bulk 2D hard-disk systems permits us to more objectively compare and discuss the properties of both. In particular, by analyzing the results of computer simulations studies~\cite{adtpre2006,HuertaJCIS2015,adtPRR2020,adtFP2021} we found that in both systems under increasing density the assemblies of hard disks are experiencing entropy-driven structural rearrangements toward the formation of the local quasiregular hexagons (in the case of bulk 2D hard disks) and the local black-red-black triangles commensurable with hexagons (in the case of a confined q1D hard disks) that is illustrated in Fig.~\ref{JCPfig2}. Upon increasing density, such structural rearrangements in each of these systems terminate by the mutual caging of a central disk: (i) by the three alternating hexagonal neighbors in the case of bulk 2D hard disks, and (ii) by the two, i.e., the left and the right neighbors in the case of confined q1D hard disks.

Based on this observation it was already argued that the caging by nearest neighbors is the origin of the freezing transition in the bulk 2D hard disks and the triangular unit cell model composed of central disk and alternating hexagonal nearest neighbors (the disks filled in blue in Fig.~\ref{JCPfig2}) has been suggested~\cite{adtpre2006}. Such a model consists of only one structural order parameter -- the distance  $\,d\,$ between alternating hexagonal nearest neighbors. Within this model, the distance $\,2\sigma\,$ is the key distance: the 2D hard disk system becomes unstable when upon the increase of density the parameter  $\,d\,$ assumes values smaller than $\,2\sigma\,$, since the gap between alternating neighbors is not large enough for the central disk to escape. It was shown~\cite{adtpre2006} by computer simulations that freezing transition occurs when the fraction of {so} caged disks in the bulk 2D hard disk system is about 40\% while the melting transition starts when it drops to below 60\%.

When such a triangular unit cell is mapped onto a q1D setup, only two (the disks filled in blue and rounded with the thick solid line in Fig.~\ref{JCPfig2}) out of three alternating hexagonal neighbors remain. However, such a reduced (due to confining walls) unit cell model still keeps unchanged the parameter $\,d\,$.  By analogy with a triangular unit cell in the case of 2D hard disks, when parameter $\,d>2\sigma\,$ the central disk is moving up and down in $\,y-$direction (the gas-like state), while in the case $\,d<2\sigma\,$ it becomes caged (the liquid-like state). Therefore, we are suggesting using this physical meaning of parameter $\,d\,$ for the definition of the structural order parameter in the case of a q1D system of hard disks. Since under q1D confinement the disk serial number remains fixed (the disk overpassing is impossible), for any disk $\,i\,$  the nearest neighbors always will be the disks $\,i-1\,$ and $\,i+1\,$ from left and right, respectively (e.g., see in Fig.~\ref{JCPfig1}), we can write down:   
\begin{equation}
{d(i)\equiv r_{\mathrm{13}}=|\mathbf{r}_{\mathrm{i+1}}-\mathbf{r}_{i-1}|}\,.
\label{di}
\end{equation}
Such a definition allows unambiguously  to track the structural order parameter on the fly in computer simulations.

\subsection*{Computer simulations}
We have conducted extensive Monte Carlo and molecular dynamics simulations of the q1D system of hard disks, both in the constant $\,NLT\,$  ensemble. In the case of  Monte Carlo (MC) simulations, the standard Metropolis technique~\cite{MetropolisJCP1953} for hard disks was used, i.e., moves are rejected if they lead to overlaps. On other hand, we employed the event driven algorithm~\cite{AlderJCP1959, DonevJComP2005} to perform molecular dynamics (MD) simulations. According to this algorithm, the temperature is kept constant by scaling appropriately the magnitude of velocities of each hard disk such that kinetic energy of the system agrees with the equipartition theorem. 

Two sizes of the system, namely, $\,N=400\,$ and  $\,N=1000\,$ hard disks have been considered. To prepare initial configurations, the chosen number $\,N\,$ of disks were placed  in the form of solid zig-zag array commensurable with 2D triangular lattice inside the channel of the length $\,L_{\rm cp}\,$ that corresponds to the close packing (cp) linear density, $\,\rho_{\rm cp}=2/\sqrt{3}=1.1547\ldots \,$ of the channel width $\,H=(3/2)\sigma\,$. Then, the initial disk configurations for the series of lower densities $\,0.8\leq\rho<\rho_{\rm cp}\,$ were obtained by gradually increasing the channel length $\,L_{\rm cp}\,$ up to the  desired length using $\,L=N\sigma/\rho\,$. The directions of the velocities in the case of MD simulations were chosen randomly. Finally, since the ends of a hard-wall channel in the present study are open, the periodic boundary conditions were applied in the $\,x$-direction.

Before  to calculate the averages, we were running both MC (over at least $\,10^8\,$ iterative steps) and MD (over at least $\,10^8\,$ collisions) 
simulations, to equilibrate the system. 


\section{Results and discussion}
The unique parameter to characterize the thermodynamic state of a q1D hard-disk system properties is the disk density. The dimensionless density parameter we are using is defined in the way similar to the case of a 1D hard-disk system as the ratio, $\,\rho = N\sigma/L\,$, and will be referred to as  linear density of the system, although the inverse of linear density -- the length per disk $\,1/\rho \equiv a =L/( N\sigma)\,$ will be used as well. In what follows, the simulation data for a set of 8 densities in a  range $\,0.8\leq\rho\leq 1.1364\,$ are discussed. For convenience the precise values of both density parameters, $\,\rho\,$ and $\,a\,$, are listed in Table~\ref{tab:table1}.

\begin{table*}[!]
\caption{\label{tab:table1} Density parameters (linear density $\rho\,$ and inverse of linear density or length per disk $\,a\,$) of the q1D system of hard disks with channel width $\,H=(3/2)\sigma\,$, that are considered in the present study.}
	\begin{ruledtabular}
		\begin{tabular}{ccccccccccc}
$\rho$&0.8&1&1.0309&1.0526&1.0870&1.1111&1.1236&0.1364\\ \hline
$\,a\,$&1.25&1&0.97&0.95&0.92&0.9&0.89&0.88 \\ 
		\end{tabular}
	\end{ruledtabular}
\end{table*}

\begin{figure}[h!]
	\centering
	\includegraphics[scale=0.60]{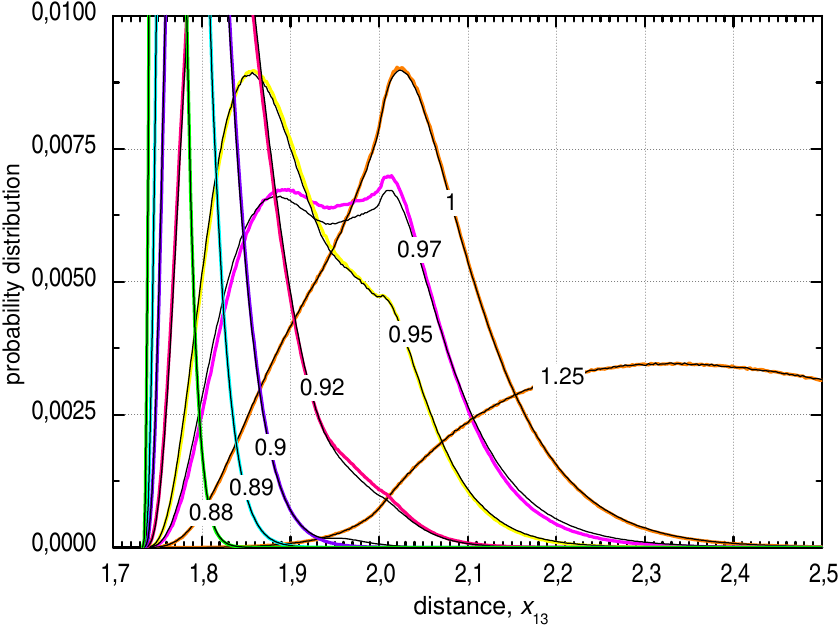}
	\caption{ MC simulation data for probability distribution {(not normalized)} of the distance {$\,d\equiv r_{13}\,$} between the left and the right nearest neighbors of the disks in q1D system composed of $\,N=400\,$ (the thick color solid line) and $\,N=1000\,$ (the thin black solid line) hard disks for the set of six thermodynamics states.	From the left to right the inverse linear densities $\,a=L/N\,$ are: {0.88 (1.754); 0.89 (1.770); 0.90 (1.787); 0.92 (1.816); 0.95 (1.854); 0.97 (1.987/2.010); 1.00 ($2.024$) and 1.25 ($\sim 2.32$) . The numbers in parenthesis are positions of the peak in corresponding probability distribution curves.}}
	\label{JCPfig3}
\end{figure}

The local structural ordering of the q1D hard-disk system can be understood in terms of the structural order parameter $\,d\,$, defined by Eq.~(\ref{di}) as {the distance $\,r\,$ between the centers of the left and right nearest neighbors of the disks. We note, that this distance always will contain exactly two hard disks -- one at the center and two halves on both sides, i.e., the quantity $\,2/x_{13}\,$ is nothing else as the local linear density of the considered system. 	Therefore, by showing in Fig.}~\ref{JCPfig3} {the probability distribution of parameter $\,d\,$  against the horizontal component $\,x_{13}\,$, we illustrate the explicit relation between the structural order parameter $\,d\,$ and the local density, that seems to be a quite natural. Although  the exact relation $\,d=\left(x_{13}^2 + y_{13}^2\right)^{1/2}\,$, where the maximal possible value of $\,y_{13} = 0.5\sigma\,$  has to be kept in mind.}

{The data presented in Fig.}~\ref{JCPfig3} { were obtained from MC simulations of the q1D system with two different sizes $\,N=400\,$ (the thick color lines) and $\,N=1000\,$ (the thin black lines) hard disks and for the full set of density parameters presented in Table}~\ref{tab:table1}. {We can see that for the case of the linear density $\,\rho =0.8\,$ (the length per disk  $\,a=1.25\,$ exceeds the disk diameter), the probability distribution of parameter $\,d\,$ is rather broad and asymmetrical. It spreads from around $\,x_{13}\approx 4\sigma\,$, with the maximum centering at the distances enough above of two disk diameters, $\,x_{13} = 2.32\sigma\,$, and down to a tiny tail in the range of  $\,x_{13}<2\sigma\,$. The latter signals that the structural order parameter $\,d\,$ might already be below the key distance $\,2\sigma\,$, pointing out an emergence of disk caging. The data in Table}~\ref{tab:table2} {confirm that indeed it is the case.}

{The shape of the probability distribution curve significantly changes as linear density increases to $\,\rho=a=1\,$. It becomes almost symmetrical concerning the peak at $\,x_{13} = 2.024\sigma\,$ that still is slightly larger than two disk diameters. It means that at this thermodynamic state, the majority of the disks ( around $\,55\%\,$ according to the data in Table}~\ref{tab:table2} )  {still are not caged. We will refer to thermodynamic states of q1D hard disks at linear densities $\,\rho \le 1\,$ as the "gas-like" although keeping in mind that caged disks are already existing.} { Almost a mirror image of the above discussed case is obtained for the probability distribution curve at linear density $\,\rho=1.0526\,$ ($\,a=0.95\,$)  with a peak position at $\,x_{13}\approx 1.85\sigma \,$. It means that at this thermodynamic state, the majority of the disks are already caged  (around $\,84\%\,$ according to the data in Table}~\ref{tab:table2} ). {We will refer to such a thermodynamic state of q1D hard disks as the "liquid-like".}

{Surprisingly, the probability distribution at the density $\,\rho=1.0309\,$ ($\,a=0.97\,$), that is in between the two previous densities, $\,\rho=1\,$ and $\,\rho=1.0526\,$, exhibits two, although lower but still well-defined peaks, at $\,x_{13}\approx 1.87\sigma\,$ and $\,2.02\sigma\,$.  This will imply for the structural order parameter $\,d\,$ the statistically significant presence of the two well-distinguished kinds of locally favored structures (and locally favored densities) in the q1D system of hard disks. } Further increase of the density from $\,\rho=1.087\,$ ($\,a=0.92\,$) to $\,\rho=1.1364\,$ ($\,a=0.88\,$) demonstrates the sharp increase of the probability distribution  {peaks, shrinking its width, and shifting the peak's position to the range of distances $\,x_{13}\,$ well shorter than two disk diameters.}

An increase in the system size from $\,N=400\,$ (the thick color lines in Fig.~\ref{JCPfig3}) to 1000 (the thin black lines Fig.~\ref{JCPfig3}) hard disks practically did not affect the probability distribution data. Indeed, data presented in this figure show that for majority of disk's densities the thick color curves and the thin black curves both follow exactly the same shape. Nevertheless, some exceptions still are well visible. The first one concerns a disk linear density $\,\rho=1.0309\,$ ($\,a=0.97\,$) when the bimodal behavior of probability distribution takes place. Therefore, at this density the system might experience an increase of fluctuations due to crossover from a "gas-like"  to a "liquid-like" state in the q1D system of hard disks.

The deviations between the thin black and thick color lines also can be noticed  for a linear density range roughly between $\,\rho=1.087\,$ ($\,a=0.92\,$) and $\,\rho=1.1236\,$ ($\,a=0.89\,$). Most probably, in this density range the q1D system is experiencing another crossover from the "liquid-like" state with non-zero values of probability distribution for distances $\,x_{13}>2\sigma\,$ to the ones with all disks being caged but still far from close packing.  We will come back to this  issue  a bit later within discussion of the decay of pair distribution function, Fig.~\ref{JCPfig5}.

 \begin{figure}[h!]
 	\centering
 	\includegraphics[width=1\linewidth]{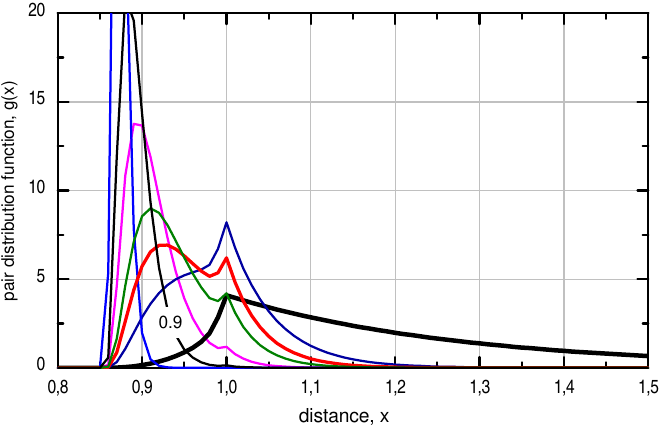}
 	\caption{ MD simulation data for pair distribution function $\,g(x)\,$ of the q1D system composed of $\,N=1000\,$ hard disks.  Figure shows the short distances that cover the first and the second coordination  shells.  The set of thermodynamics states characterized by inverse linear densities from the left to right includes:  $\,a=L/N =\,$ 0.88; 0.9; 0.92; 0.95; 0.97; 1 and 1.25.  } 
 	\label{JCPfig4}
 \end{figure}

The {peak's positions in probability distribution of the structural order parameter $\,d\,$ determines the size of the first coordination shell of the disks (see  the data in parenthesis behind of each density in the caption to Fig.}~\ref{JCPfig3}).  {This information usually is estimated from the first minimum's position in the pair (or radial) distribution function and is referred to as the radius of the first coordination shell. Such a function is rather well studied for the q1D system of hard disks}~\cite{Varga2011,Varga2013,adtPRR2020,adtFP2021,Montero2023}. {Figure}~\ref{JCPfig4} {shows data for pair distribution function $\,g(x)\,$ of the q1D system of hard disks that were obtained in present study from MD computer simulations of larger ($\,N=1000\,$) system.  It presents the short distances that cover the first coordination  shell only.  The thermodynamic states are the same as for probability distribution in Fig.}~\ref{JCPfig3}.  
By comparing sets of data presented on Figs.~\ref{JCPfig3} and \ref{JCPfig4} we found that from $\,g(x)\,$ alone not always is possible to extract the quantitative information on structural order parameter $\,d\,$.  At least for the case of considered q1D system of hard disks. 

 {However, the important part of  qualitative information obtained from the probability distribution of the structural order parameter $\,d\,$  still is contained in the pair distribution function $\,g(x)\,$. By this information we mean the split of a first peak of $\,g(x)\,$ at linear density  $\,\rho=1.0309\,$ ($\,a=0.97\,$) that in fact is indication of the presence of  local structure bimodality. However, we are not aware of such an interpretation of the phenomenon of peak's splitting in the pair distribution function $\,g(x)\,$  in the literature. Including both the q1D and 2D systems of hard disks. }

There is another important piece of information that concerns the bimodality shown in Fig.~\ref{JCPfig3} and the pair distribution function $\,g(x)\,$ already reported in the literature. It is about the appearance  of unexpected oscillating tails  in the behavior of $\,g(x)-1\,$ at  long distances, having the magnitude dependent on linear density up to about  $\,10^{-1}\,$  (e.g., see Fig.~2b in Ref.~\cite{HuPRR2021} and Fig.~4b in Ref.~\cite{adtJML2023}). Since such oscillating tails usually are considered as a  kind of noise, authors usually did not show this part of simulation data by cutting $\,y$-axes in log-log plots at the smallest at $\,10^{-1}\,$. 
Figure~\ref{JCPfig5} {presents  our simulations data for $\,g(x)-1\,$ with $\,y$-axes down to $\,10^{-3}\,$ and in the range of linear densities from $\,\rho=0.8\,$  ($\,a=1.25\,$) up to $\,\rho=1.1364\,$  ($\,a=0.88\,$).

		\begin{figure}[h!]
			\includegraphics[width=0.95\linewidth]{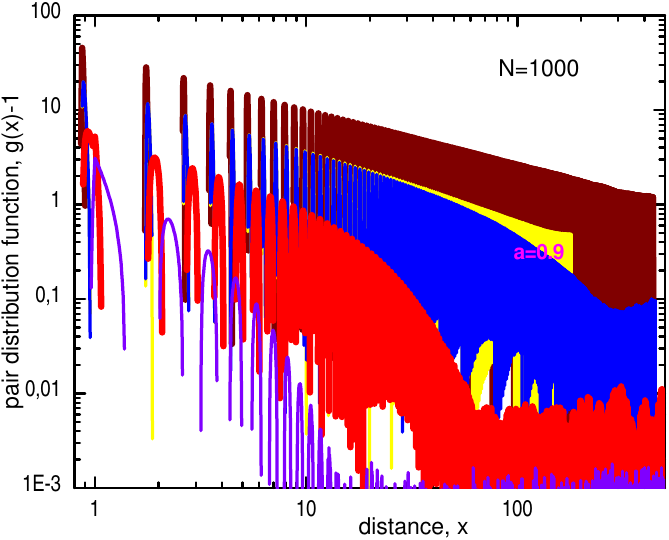}
			\caption{ Log-log plot of the pair distribution function $\,g(x)-1\,$ of the q1D system composed of $\,N=1000\,$ hard disks at the particular set of thermodynamics states determined by the inverse linear densities $\,a=L/N =\,$ 1.25; 0.97; 0.9 and 0.88 (from the bottom to the top). For the case of $\,a=0.9\,$ the data for $\,N=400\,$ are shown as well.}
			\label{JCPfig5}
		\end{figure}

{The oscillations we are talking about already are seen at the lowest considered linear density $\,\rho=0.8\,$ and are growing as density $\,\rho\,$ increases. We argue that such a "noise" reflects the long-range correlations originated by the presence in the considered q1D system of the density-dependent amount of caged disks. Implicitly this is evidenced by a non-zero probability distribution values of the structural order parameter $\,d\,$ for short distances $\,x_{13}<2\sigma\,$ in Fig.}~\ref{JCPfig3}, {and is confirmed explicitly  by the data collected in Table}~\ref{tab:table2}.  {This table consists of the set of averaged values for q1D system generated in several consecutive runs for all densities considered. The averaged quantities include the fraction of caged disks, the distance $\,r_{13}\,$, and its shorter $\, r_{13}<2\,$ and wider $\, r_{13}>2\sigma\,$ realizations.} {Additionally, in Fig.}~\ref{JCPfig6} {we present computer simulation data showing the formation of the clusters of caged disks of varied length and concentration. It follows that the q1D system of hard disks in fact is a mixture of uncaged "gas-like"  and caged "liquid-like" disks. Being of the short length, the caged disk clusters result in the "noise" in the disk-disk pair distribution function (the case of low density $\,\rho=0.125\,$ in Fig.}~\ref{JCPfig5}).

\begin{figure}[h!]
	\includegraphics[width=0.975\linewidth]{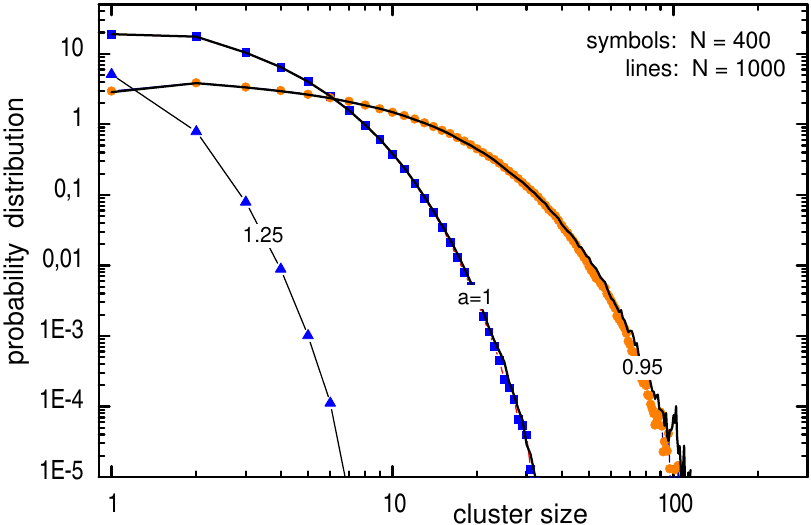}
	\caption{{MD simulation data for probability distribution (per number $\,N\,$ of particles) of the size of clusters formed by caged disks in q1D system of hard disks upon increase of the density from $\,\rho=0.8\,$ ($\,a=1.25\,$)  to $\,1.0526\,$ ($\,a=0.95\,$). The data for the lowest density are the first, while for the highest density are the last from left to right at the bottom of the figure. The symbols correspond to the system of $\,N=400\,$ while the lines correspond to the system of $\,N=1000\,$.  The single cluster (cluster of size 1) consists of one disk caged by its left and right neighbors (three disks in total); the cluster of size 2 consists of two caged disks (four disks in total), and so on.}
	}
	\label{JCPfig6}
\end{figure}

{Upon increasing linear density, the probability distribution peaks in Fig}.~\ref{JCPfig3} { are growing and curves are shifting to shorter distances $\,x_{13}\,$,  increasing the amount of caged disks in the system and strengthening of the overall disk-disk long-range correlations including those responsible for oscillating tail (e.g., see the case of  inverse density $\,a=0.97\,$  in Fig.}~\ref{JCPfig5}).  At the same time,  increase of the fraction of caged disks by increasing the disk's density leads to decrease of the amount of single clusters, while longer clusters continue growing (e.g., see the case of  inverse density $\,a=0.95\,$  in Fig.~\ref{JCPfig6} ). This manifests itself in the emergence of the power-law decay of the disk-disk pair correlations at short distances (e.g., see the case of inverse densities $\,a=0.97\,$ and 0.9 in Fig.}~\ref{JCPfig5}). 
	
{On the other hand, the high degree of disks caging in the inverse density range $\,0.95>a>0.9\,$ culminates in the non-zero value of the probability distribution of disk's complete clustering (does not shown in Fig.}~\ref{JCPfig6})  at inverse density $\,a\le 0.91\,$. Although the windows, $\,\langle r_{13}\rangle >2\sigma\,$, according to the data in Table~\ref{tab:table2} { also still exist. However, in contrast to the case of  $\,a > 0.91\,$, at smaller inverse densities,  those windows represent the space between the ends of non-completed clusters. The shape of the decay of long-range correlations at the density that is still before the complete clustering is well illustrated by the case of $\,a=0.9\,$  in Fig.}~\ref{JCPfig5}.
{This density seems to be  particular for the considered q1D system of hard disks. The well-defined minimum on the shape of the correlation decay of the pair distribution function $\,g(x)-1\,$, although approximate, indicates the largest possible size ( $\,\sim 300\,$ caged disks) of the "liquid-like" clusters in a q1D system of hard disks. According to such an estimate, the systems of the size smaller than $\,N=600\,$ most probably might be sensitive to the system-size effects. Indeed, it is the case we are observing in  Fig.}~\ref{JCPfig5} {for the case of the system size $\,N=400\,$.}

\begin{table}[!]
	\caption{\label{tab:table2} MD computer simulation data for averaged value of the structural order parameter  $\,\langle d \rangle = \langle r_{13} \rangle\,$ and averaged value of the fraction of disks that are caged. For the former, it is also shown separately the contributions from both sides of the key distance $\,2\sigma\,$, i.e., $\,r_{13}<2\sigma\,$ and $\,r_{13}>2\sigma\,$. For each density, the first row corresponds to the system-size of $\,N=1000\,$ disks, while the row below shows the data for $\,N=400\,$ disks.}
	\begin{ruledtabular}
		\begin{tabular}{ccccc}
			inverse&total &smaller&larger&fraction\\
			density&distance&distance&distance&of        \\
			$a=1/\rho$&$\langle r_{13}\rangle$&$\langle r_{13}<2\sigma\rangle$&$\langle r_{13}>2\sigma\rangle$& caged\footnote{the disks for which parameter $d=r_{13}<2\sigma$}  \\
			\hline
			1.25 &2.5090 &1.9585& 2.5186 &0.01719(2) \\
			&2.5090 &1.9588& 2.5188 &0.01731(0)\\
			1.00 &2.0114 &1.9229& 2.0849 &0.45327(7)\\
			&2.0115 &1.9229& 2.0849 &0.45303(9)\\
			0.97 &1.9494 &1.8999& 2.0616 &0.69410(7)\\ 
			&1.9494 &1.9000& 2.0615 &0.69434(0)\\
			0.95 &1.9069 &1.8795& 2.0500 &0.83958(5)\\
			&1.9069 &1.8796& 2.0498 &0.83981(8)\\
			0.92 &1.8421 &1.8376& 2.0342 &0.97722(2)\\
			&1.8422&1.8376 & 2.0339 &0.97686(23)\\
			0.90&1.8001 &1.8001& 2.0182 &0.99993(0)\\
			&1.8001 &1.8001& 2.0186(2)&0.99993(0) \\
			0.89&1.7800 &1.7800& 2.0124(7)&0.99999(0) \\
			&1.7800 &1.7800& 2.0089(12)&0.99999(0) \\
			0.88&1.7600 &1.7600& --        &1              \\ 
			&1.7600 &1.7600& --         &1              \\ 
		\end{tabular} 
	\end{ruledtabular}   
\end{table}

{As soon as probability distribution curves in Fig.}~\ref{JCPfig3} {are shrinking to exhibit  the sharp single maximum (e.g., see the curve for $\,a=0.88\,$), the correlations in Fig.}~\ref{JCPfig5} {demonstrate the tendency to follow the power-law-like decay, the behavior that serves to identify the Kosterlitz-Thouless-type transition}~\cite{KT}.

The possibility of the Kosterlitz-Thouless-type transition in a q1D system under study using above criteria has been already suggested based on computer simulations of $\,N=400\,$ hard disks~\cite{adtPRR2020}. In that study a characteristic power law decay that persists up to $\,x\sim 200\,$ was reported for linear density  $\,\rho=1.1111\,$ ($\,a=0.9\,$). {According to above discussion of the probability distribution data in Fig.}~\ref{JCPfig3}, {the density $\,\rho=1.1111\,$ ($\,a=0.9\,$) falls within the range of densities for which the properties may depend on the system size.} The data presented in Fig.~\ref{JCPfig5} demonstrate that this is the case. Indeed,  for simulation data obtained with $\,N=1000\,$ (the blue lines in Fig.~\ref{JCPfig5}) the power-law-like decay starts to decline already at $\,x\sim 50\,$. It is also in line with Hu and Charbonneau~\cite{HuPRR2021} who  using the results obtained from planting scheme applied to q1D system of hard disks under transfer matrix solution, already observed that for this  density the power-law-like decay terminates at $\,x\sim 100\,$.

As it would be expected from the probability distribution in Fig.~\ref{JCPfig3}, the further increase in density restores the power-law-like decay. To illustrate this, in Fig.~\ref{JCPfig5} we present the data (the wine color lines) for linear density $\,\rho=1.1364\,$ ($\,a=0.88\,$) that demonstrate a power-law-like decay persisting at least up to $\,x\sim 500\,$ (the maximal distance for q1D system with $\,N=1000\,$ hard-disk particles).

\section{Conclusions}

The essential difficulty in defining locally favored states in the case of complex fluids is in finding a proper structural order parameter~\cite{RussoNC2014}. Since the geometry of quasi-1D confinement does not allow disks to pass each other, the first coordination shell of disk $\,i\,$ is formed just by its two neighboring disks, namely, $\,i-1\,$ and $\,i+1\,$ (see Fig.~\ref{JCPfig1}a).  Therefore q1D hard disks considered in the present study are the simplest non-trivial system that allows unambiguous definition of the structural order parameter in the form of Eq.~\ref{di}. 

The observed bimodal probability distribution, Fig.~\ref{JCPfig3}, implies that the considered q1D system of hard disks at linear density $\,\rho=1.0309\,$ ($\,a=0.97\,$) is a mixture of two statistically significant local structures and local densities characterized by different values of the structural order parameter, $\,d=1.987\sigma\,$ (the "liquid-like") and $\,2.010\sigma\,$ (the "gas-like"), i.e., being shorter and longer than the key distance $\,2\sigma\,$ between the left and the right neighbors of disks in the system. 

{The suggested structural order parameter $\,d\,$ allows to explicitly identify the origin of size dependence of the length scale over which the power law in pair distribution function $\,g(x)-1\,$ extends. Such an origin relies on the existence of the system size-dependent "tail" in probability distribution of parameter $\,d\,$, and that "tail" concerns the range of $\,d\,$ values in the vicinity of $\,2\sigma\,$. This finding is illustrated in Fig.}~\ref{JCPfig3} {for the linear density $\,\rho=1.1111\,$ $\,(a=0.9)\,$. For slightly higher densities $\,\rho=1.1236\,$ $\,(a=0.89)\,$ and 1.1364 $\,(a=0.88)\,$ the peaks of probability distribution of parameter $\,d\,$ become sharper by shifting entire distribution curves to shorter distances $\,x_{13}\,$, while the system size-dependent "tails" vanish. The latter occurs in the same way for both system sizes considered and no system size-dependency is observed at least for linear densities $\,\rho>1.1364\,$ $\,(a=0.88)\,$.}

Finally, based on the unit cell model definition of the structural order parameter $\,d\equiv r_{13}\,$ and simulations data for the probability distribution of this parameter in Fig.~\ref{JCPfig3}, in conjunction with simulation data for pair distribution function $\,g(x)\,$, the present study shows that power-law-like decay of pair correlations in q1D hard disks has its origin in 2D caging physics.

\begin{acknowledgments}
    The authors are indebted to V.M. Pergamenshchik for numerous discussions.
	AT acknowledge financial support received through the MSCA4Ukraine project, which is funded by the European Union, grant agreement ID: 1039539.
	AH thankfully acknowledge computer resources, technical advice and support provided by Laboratorio Nacional de Supercomputo del Sureste de Mexico (LNS), a member of the CONACYT national laboratories, with project No. 202203061N. AH also thanks the support of CONACyT, Mexico during the sabbatical leave.
	
\section*{Data Availability Statement}
The data that support the findings of this study are available from the corresponding author upon reasonable request.
	
\end{acknowledgments}


\end{document}